\documentstyle[12pt,axodraw]{article}

\begin{document}
\baselineskip 0.7cm

\newcommand{\gsim}{ \mathop{}_{\textstyle \sim}^{\textstyle >} }
\newcommand{\lsim}{ \mathop{}_{\textstyle \sim}^{\textstyle <} }
\newcommand{\EV}{ {\rm eV} }
\newcommand{\KEV}{ {\rm keV} }
\newcommand{\MEV}{ {\rm MeV} }
\newcommand{\GEV}{ {\rm GeV} }
\newcommand{\TEV}{ {\rm TeV} }
\renewcommand{\thefootnote}{\fnsymbol{footnote}}
\setcounter{footnote}{1}

\begin{titlepage}
%\today
\begin{flushright}
UT-797, OHSTPY-HEP-T-97-020
\\
Nov, 1997
\end{flushright}

\vskip 0.35cm
\begin{center}
{\large \bf 
A Gauge Mediation Model of Dynamical Supersymmetry Breaking
without Color Instability}
\vskip 1.2cm
Yasunori~Nomura$^{a)}$, K.~Tobe$^{b)}$, and T.~Yanagida$^{a)}$
\vskip 0.4cm

a) {\it Department of Physics, University of Tokyo,\\
     Bunkyo-ku, Hongou, Tokyo 113, Japan}
\\
b) {\it Department of Physics, The Ohio State University,\\
     Columbus, Ohio 43210, USA}
\vskip 1.5cm

\abstract{We construct a gauge mediation model of dynamical 
supersymmetry breaking (DSB) based on a vector-like gauge theory, 
in which there is a unique color-preserving true vacuum.
The DSB scale $\Lambda/4\pi$ turns out to be as high as 
$\Lambda/4\pi \simeq 10^{8-9}~\rm{GeV}$, since the transmission 
of the DSB effects to the standard model sector is completed 
through much higher loops.
This model is perfectly natural and phenomenologically consistent.
We also stress that the dangerous D-term problem for the messenger 
U(1)$_m$ is automatically solved by a charge conjugation symmetry 
in the vector-like gauge theory.}
\end{center}
\end{titlepage}

\renewcommand{\thefootnote}{\arabic{footnote}}
\setcounter{footnote}{0}

%
%
%       *** Main Part ***
%
%

In the last couple of years there has been a growing interest 
in dynamical supersymmetry breaking (DSB) with gauge mediation \cite{DFS}, 
since it provides natural solutions to various phenomenological 
problems such as flavor-changing neutral current problem 
and large CP violation.
Although many realistic models for gauge mediation have been 
constructed \cite{DNS,DNNS,HIY,INTY,PT,HMM}, no satisfactory and simple model 
has been found yet.

Among various models proposed so far the original model 
with messenger quarks and leptons in Ref.~\cite{DNS,DNNS} seems 
the most attractive and natural.
However, this original model has color-breaking global minima \cite{DDR} 
and hence one needs to postulate that our universe sits at 
a local minimum preserving the color SU(3).

The purpose of this letter is to improve the original model 
so that the true vacuum indeed preserves the color SU(3).
In the present model we keep all attractive features 
in the original model \cite{DNS,DNNS}.
The present model is perfectly natural and phenomenologically consistent.
We also point out that a serious D-term problem for the 
messenger gauge interaction is automatically solved 
by a natural symmetry exists in the present model.

The model is based on DSB in vector-like gauge theories 
observed by Izawa and one of the authors (T.Y.) \cite{IY} 
and Intriligator and Thomas \cite{IT}.
We adopt here a supersymmetric (SUSY) SU(2) gauge theory 
with four doublet chiral superfields 
$Q_i$ and six singlet ones $Z^{ij} = -Z^{ji}$, 
where $i$ and $j$ are flavor indices ($i,j=1,\cdots,4$). 
This theory has a flavor SU(4)$_F$ symmetry.
We introduce the following tree-level superpotential 
to break explicitly the unwanted global SU(4)$_F$ ; 
\begin{eqnarray}
 W_{\rm tree} = \lambda_{ij} Z^{ij} (Q_i Q_j).
\end{eqnarray}
To make our point clear we impose a global SP(4)$_F$ 
and then the superpotential is written in a simpler form as 
\begin{eqnarray}
 W_{\rm tree} = \lambda Z (QQ) + \lambda_Z Z^a (QQ)_a,
\end{eqnarray}
where $(QQ)$ and $Z$ are singlets of SP(4)$_F$ and $(QQ)_a$ and $Z^a$ are 
five-dimensional representations of SP(4)$_F$.

The low energy effective superpotential
may be written \cite{IS} in terms of gauge-invariant 
low-energy degrees of freedom
\begin{eqnarray}
 V \sim \frac{4\pi}{\Lambda} (QQ), \quad 
 V_a \sim \frac{4\pi}{\Lambda} (QQ)_a
\end{eqnarray}
as follows:
\begin{eqnarray}
 W_{\rm eff} = X(V^2 + V_a^2 - \frac{\Lambda^2}{(4\pi)^2}) 
           + \frac{\lambda}{4\pi} \Lambda Z V 
           + \frac{\lambda_Z}{4\pi} \Lambda Z^a V_a,
\end{eqnarray}
where $X$ is an additional chiral superfield and $\Lambda$ 
a dynamical scale of the SU(2) gauge interaction.
Factors of $4\pi$ are added by using a na\"{\i}ve dimensional analysis 
(NDA) \cite{NDA}.
For a relatively large value of the coupling $\lambda_Z$, 
we obtain the SP(4)$_F$ invariant condensation,
\begin{eqnarray}
 \langle V \rangle = \frac{\Lambda}{4\pi}, 
\label{condensation}
\end{eqnarray}
which implies 
\begin{eqnarray}
 \langle (QQ) \rangle \equiv 
     \langle \frac12 (Q_1 Q_2 + Q_3 Q_4) \rangle 
     \sim (\frac{\Lambda}{4\pi})^2.
\label{condensation2}
\end{eqnarray}
Then, the low-energy effective superpotential is approximately given by
\begin{eqnarray}
 W_{\rm eff} \simeq \frac{\lambda}{(4\pi)^2} \Lambda^2 Z.
\end{eqnarray}
We have nonvanishing $F$ term, $\langle F_Z \rangle 
\simeq \lambda \Lambda^2 / (4\pi)^2 \neq 0$, and 
hence SUSY is broken \cite{IY,IT}.

We gauge a U(1)$_m$ subgroup in the SP(4)$_F$.\footnote{
Gauging the U(1)$_m$ symmetry does not restore SUSY.}
The charge assignments of chiral superfields under U(1)$_m$ are 
assumed to be 
\begin{equation}
\begin{array}{c}
Q_{1} (+1), \quad Q_{2} (-1), \quad Q_{3} (0), \quad Q_{4} (0),\nonumber\\
\\
Z^{12} (0), \quad Z^{34} (0), \quad Z^{13} (-1), \quad 
Z^{14} (-1), \quad Z^{23} (+1), \quad Z^{24} (+1),
\label{charge_assignment}
\end{array}
\end{equation}
where the numbers in each parentheses denote the U(1)$_m$ charges.
Note that the above condensation Eq.~(\ref{condensation}) does not 
break the U(1)$_m$ gauge symmetry, since $V$ is neutral under the 
U(1)$_m$.\footnote{
We assume the minimal K\"{a}hler potential for the $Z$ field for simplicity.
$Z$ may get a non-zero vacuum expectation value if one includes higher-order 
corrections to the K\"{a}hler potential, since the $Z$ direction is 
flat at the tree level.
Even if it is the case, however, the U(1)$_m$ is not broken 
since $Z$ has also vanishing charge of the U(1)$_m$, 
and our main conclusion is unchanged.}

Here, we comment on the light degrees of freedom in the DSB sector in the 
present model. The present model has a larger global symmetry SP(4)$_F
\times$ U(1)$_{\chi} \times$ U(1)$_R$ and the U(1)$_{\chi}$ is
spontaneously broken by the condensation of
Eq.~(\ref{condensation2}). However, there appears no massless 
Nambu-Goldstone boson, since the U(1)$_{\chi}$ has SU(2) gauge
anomaly. The 't Hooft's anomaly matching condition for the remaining 
SP(4)$_F$ and the nonanomalous U(1)$_R$ is satisfied by the $Z$ field. 
Thus, the chiral superfield $Z$ is only the flat direction in the
present model. This fact is a crucial feature different from the original
model~\cite{DNS,DNNS} which contains light fields with nonvanishing 
U(1)$_m$ charges in the DSB sector.

We now consider the messenger sector.
The messenger fields are nothing but those in Ref.~\cite{DNS,DNNS} : 
we have three chiral superfields,
\begin{eqnarray}
E (+1), \quad \bar{E} (-1), \quad S (0),
\end{eqnarray}
and vector-like messenger quark and lepton superfields, 
$d$, $\bar{d}$, $l$, $\bar{l}$.
Here, the messenger quark multiplets 
$d$, $\bar{d}$ and lepton multiplets $l$, $\bar{l}$ 
are all neutral under U(1)$_m$.
We assume that the $d$ and $\bar{d}$ transform as 
the right-handed down quark and its antiparticle
under the standard model (SM) gauge group, respectively.  
The $l$ and $\bar{l}$ are assumed to transform as 
the left-handed lepton doublet and its antiparticle, respectively.
The most general superpotential for the messenger sector 
without any dimensional parameters is 
\begin{eqnarray}
 W_{\rm mess} = k_E S E \bar{E} + \frac{f}{3} S^3 
            + k_d S d \bar{d} + k_l S l \bar{l}.
\label{superpotential}
\end{eqnarray}
This is the same form as in the original model \cite{DNS,DNNS} 
and natural in the sense that there is a consistent symmetry 
which forbids other terms in the superpotential.

When we switch on the U(1)$_m$ gauge interaction, 
the soft SUSY-breaking masses for 
$E$ and $\bar{E}$ fields are generated through the SUSY-breaking 
effects in the DSB sector.
The resulting scalar potential is 
\begin{eqnarray}
 V_{\rm mess} = \sum_\eta |\frac{\partial W}{\partial \eta}|^2 
                   + m_E^2 |E|^2 + m_{\bar{E}}^2 |\bar{E}|^2,
\label{scalar_potential}
\end{eqnarray}
where $m_E$ and $m_{\bar{E}}$ are soft SUSY-breaking masses for 
$E$ and $\bar{E}$, and 
$\eta$ denotes chiral superfields 
$E, \bar{E}, S, d, \bar{d}, l$, and $\bar{l}$.
One of the essential differences between the present model 
and the original one is that the soft SUSY-breaking mass squared 
$m_E^2$ and $m_{\bar{E}}^2$ are positive in the present model 
while these are negative in the original model.
We will discuss this point below.  

There are two types of quantum corrections 
to the soft SUSY-breaking masses for the messenger fields: 
one is from 'threshold effects' and the other from 'running effects'.
In Ref.~\cite{DNS,DNNS}, the $m_E^2$ and $m_{\bar{E}}^2$ come 
essentially from the 'running effects'.
The DSB sector contains light superfields $\chi^{\pm}$ 
in the original model which have U(1)$_m$ charges.
Since both of SUSY-breaking soft masses and the SUSY-invariant mass
for the superfields $\chi^{\pm}$ are very small 
compared with the DSB scale $\Lambda$, 
the 'running effects' from the SUSY-breaking scale to the $\chi$ 
field mass scale becomes important.
According to the two-loop renormalization group equation (RGE), 
we find that the $m_E^2$ and $m_{\bar{E}}^2$ 
receive negative contributions from the $\chi$ 
and gauge multiplet loops, 
which results in negative $m_E^2$ and $m_{\bar{E}}^2$.
With these negative SUSY-breaking masses the DSB effects are
transmitted to the messenger quarks and leptons, 
which causes, however, instability of the color-preserving 
vacua \cite{DDR}.

On the other hand, the 'threshold effects' dominate over the 
'running effects' in the present model.
The reason is that since the DSB sector is a vector-like theory 
and the flavor SP(4)$_F$ is unbroken, all composite fields have
dynamical masses of order $\Lambda$ and there are no light field besides
the $Z$ superfield as discussed above.
%The present model has, in fact, a larger global
%symmetry SP(4)$_F \times$ U(1)$_{\chi} \times$ U(1)$_R$ and the
%U(1)$_{\chi}$ is spontaneously broken by the condensation of
%Eq.~(\ref{condensation2}). However, there appears no massless
%Nambu-Goldstone boson, since the U(1)$_{\chi}$ has SU(2) gauge
%anomaly. The 't Hooft's anomaly matching condition for the remaining 
%SP(4)$_F$ and the nonanomalous U(1)$_R$ is satisfied by the $Z$ field. 
%Thus, the chiral superfield $Z$ is only the flat direction in the
%present model. 
Therefore, the 'running effects' which would drive 
the $m_E^2$ and $m_{\bar{E}}^2$ negative is 
negligible compared with relatively large 'threshold effects'.
As a consequence we have positive SUSY-breaking masses for the 
$E$ and $\bar{E}$ fields. 

Let us evaluate the threshold corrections to the $m_E^2$ and
$m_{\bar{E}}^2$. It may be very difficult to 
calculate them directly, because of the strong gauge dynamics 
in the SUSY breaking sector. 
Therefore, we estimate them by using the effective theory resulting from the
strong gauge dynamics. The effective theory contains many composite fields 
which consist of the elementary $Q$ fields. Here, 
we assume that the lightest composite fields are 
a pair of chiral superfields $\Phi$ and $\bar{\Phi}$ which 
have U(1)$_m$ charges $\pm 1$, respectively.\footnote{
In fact, it has been shown that the lightest bound state of 
the SUSY QED consists of a pair of chiral supermultiplets 
in Ref.~\cite{BLP}. This fact may support our assumption.}
We also consider that they give a dominant contribution to 
the masses $m_E^2$ and $m_{\bar{E}}^2$.

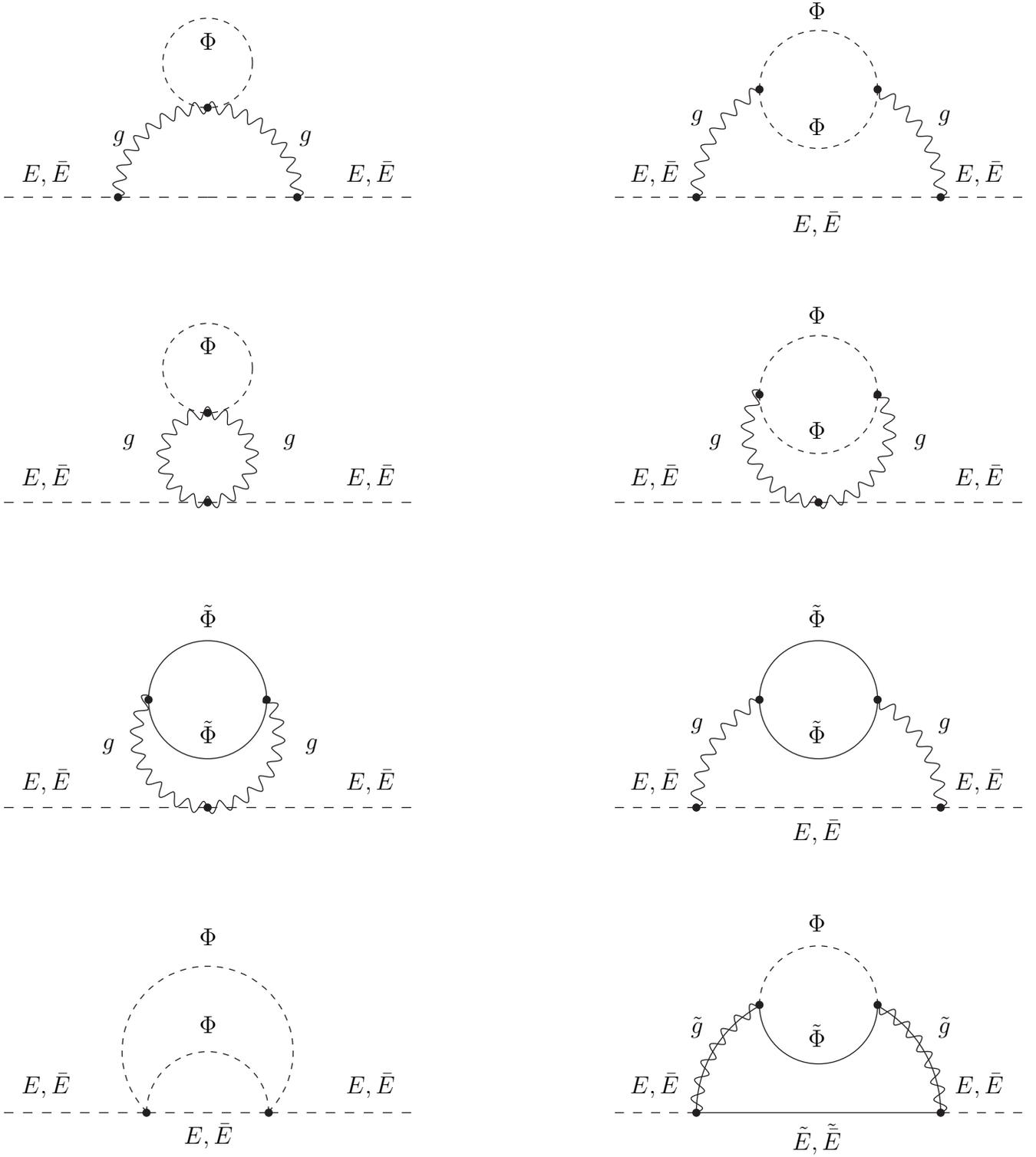
\begin{figure}
\begin{center} 
\begin{picture}(650,540)(-150,120)
% Graph 1
  \DashLine(-190,600)(-90,600){5} \Text(-170,610)[b]{$E,\bar{E}$}
  \Vertex(-134,600){2} \Vertex(-46,600){2}
  \Text(-131,630)[rb]{$g$} \PhotonArc(-90,600)(44,0,180){3}{17}
  \Vertex(-90,644){2} \Text(-45,630)[lb]{$g$}
  \DashCArc(-90,666)(22,0,360){3} \Text(-90,675)[b]{$\Phi$}
  \DashLine(-90,600)(10,600){5} \Text(-10,610)[b]{$E,\bar{E}$}

% Graph 3
  \DashLine(-190,450)(-90,450){5} \Text(-170,460)[b]{$E,\bar{E}$}
  \Vertex(-90,450){2} \Text(-126,480)[rb]{$g$}
  \PhotonArc(-90,472)(22,0,360){3}{17} \Vertex(-90,494){2}
  \Text(-53,480)[lb]{$g$} \DashCArc(-90,516)(22,0,360){3}
  \Text(-90,525)[b]{$\Phi$} \DashLine(-90,450)(10,450){5}
  \Text(-10,460)[b]{$E,\bar{E}$}

% Graph 5
  \DashLine(-190,300)(-90,300){5} \Text(-170,310)[b]{$E,\bar{E}$}
  \Vertex(-90,300){2} \PhotonArc(-90,335)(35,145,35){3}{18}
  \Text(-136,330)[rb]{$g$} \Vertex(-119,353){2}
  \Text(-42,330)[lb]{$g$} \Vertex(-61,353){2}
  \CArc(-90,353)(29,0,180) \Text(-90,390)[b]{$\tilde{\Phi}$}
  \CArc(-90,353)(29,180,360) \Text(-90,333)[b]{$\tilde{\Phi}$}
  \DashLine(-90,300)(10,300){5} \Text(-10,310)[b]{$E,\bar{E}$}

% Graph 7
  \DashLine(-190,150)(-120,150){5} \Text(-170,160)[b]{$E,\bar{E}$}
  \Vertex(-120,150){2} \DashLine(-120,150)(-60,150){5}
  \Text(-90,145)[t]{$E,\bar{E}$} \DashCArc(-90,150)(30,0,180){3}
  \DashCArc(-90,180)(42,315,225){3}
  \Text(-90,233)[b]{$\Phi$}
  \Text(-90,190)[b]{$\Phi$} \DashLine(-60,150)(10,150){5}
  \Text(-10,160)[b]{$E,\bar{E}$} \Vertex(-60,150){2}

% Graph 2
  \DashLine(110,600)(150,600){5} \Text(130,610)[b]{$E,\bar{E}$}
  \Vertex(150,600){2} \DashLine(150,600)(270,600){5}
  \Text(210,595)[t]{$E,\bar{E}$} \Text(154,640)[rb]{$g$}
  \PhotonArc(210,600)(60,120,180){3}{7} \Vertex(181,653){2}
  \Text(270,640)[lb]{$g$} \PhotonArc(210,600)(60,0,60){3}{7}
  \Vertex(239,653){2} \DashCArc(210,653)(29,0,180){3}
  \Text(210,690)[b]{$\Phi$}
  \DashCArc(210,653)(29,180,360){3}
  \Text(210,633)[b]{$\Phi$} \DashLine(270,600)(310,600){5}
  \Text(290,610)[b]{$E,\bar{E}$} \Vertex(270,600){2}

% Graph 4
  \DashLine(110,450)(210,450){5} \Text(130,460)[b]{$E,\bar{E}$}
  \Vertex(210,450){2} \PhotonArc(210,485)(35,145,35){3}{18}
  \Text(163,480)[rb]{$g$} \Vertex(181,503){2}
  \Text(258,480)[lb]{$g$} \Vertex(239,503){2}
  \DashCArc(210,503)(29,0,180){3} \Text(210,540)[b]{$\Phi$}
  \DashCArc(210,503)(29,180,360){3}
  \Text(210,483)[b]{$\Phi$} \DashLine(210,450)(310,450){5}
  \Text(290,460)[b]{$E,\bar{E}$}

% Graph 6
  \DashLine(110,300)(150,300){5} \Text(130,310)[b]{$E,\bar{E}$}
  \Vertex(150,300){2} \DashLine(150,300)(270,300){5}
  \Text(210,295)[t]{$E,\bar{E}$} \Text(154,340)[rb]{$g$}
  \PhotonArc(210,300)(60,120,180){3}{7} \Vertex(181,353){2}
  \Text(270,340)[lb]{$g$} \PhotonArc(210,300)(60,0,60){3}{7}
  \Vertex(239,353){2} \CArc(210,353)(29,0,180)
  \Text(210,390)[b]{$\tilde{\Phi}$} \CArc(210,353)(29,180,360)
  \Text(210,333)[b]{$\tilde{\Phi}$} \DashLine(270,300)(310,300){5}
  \Text(290,310)[b]{$E,\bar{E}$} \Vertex(270,300){2}

% Graph 8
  \DashLine(110,150)(150,150){5} \Text(130,160)[b]{$E,\bar{E}$}
  \Vertex(150,150){2} \Line(150,150)(270,150) 
  \Text(210,145)[t]{$\tilde{E},\tilde{\bar{E}}$}
  \CArc(210,150)(60,120,180) \Text(154,190)[rb]{$\tilde{g}$}
  \PhotonArc(210,150)(60,120,180){3}{7} \Vertex(181,203){2}
  \CArc(210,150)(60,0,60) \Text(270,190)[lb]{$\tilde{g}$}
  \PhotonArc(210,150)(60,0,60){3}{7} \Vertex(239,203){2}
  \DashCArc(210,203)(29,0,180){3} \Text(210,240)[b]{$\Phi$}
  \CArc(210,203)(29,180,360) \Text(210,183)[b]{$\tilde{\Phi}$}
  \DashLine(270,150)(310,150){5} \Text(290,160)[b]{$E,\bar{E}$}
  \Vertex(270,150){2}

%\put(37,80){\Huge{Fig.1}}
\end{picture}
\caption{Diagrams contributing to masses of the scalar components of 
$E$ and $\bar{E}$ chiral supermultiplets.
Here, $g$ and $\tilde{g}$ represent U(1)$_m$ gauge boson and gaugino, 
respectively. 
Scalar and fermion components of 
the chiral supermultiplets are denoted by the same symbols of 
the chiral superfields and those with tilde, respectively.} 
\label{messenger_soft_mass}
\end{center}
\end{figure}

Following the NDA prescription in Ref.~\cite{NDA}, the effective theory
concerning $\Phi$, $\bar{\Phi}$, $Z$ fields
is described by
\begin{eqnarray}
 {\cal L} = \frac{\Lambda^4}{g_2^2} \left[ 
 \int \frac{d^4 \theta}{\Lambda^2} \tilde{K}(\phi,\phi^{\dagger},
 \bar{\phi},\bar{\phi}^{\dagger}) +
 \int \frac{d^2 \theta}{\Lambda} \tilde{W}(\phi,\bar{\phi},z) 
 + {\rm h.c.} \right],
\end{eqnarray}
where $g_2$ is a strong SU(2) gauge coupling ($g_2 \simeq 4 \pi$)
and $\tilde{W}$ is a function of dimensionless superfields
$\phi=g_2\Phi/\Lambda$, $\bar{\phi}=g_2\bar{\Phi}/\Lambda$, 
and $z=\lambda Z/\Lambda$ as
\begin{eqnarray}
 \tilde{W} \simeq \phi \bar{\phi}+ z \phi \bar{\phi},
\
\label{NDA_W}
\end{eqnarray}
where we have assumed that all coupling constants of Eq.(\ref{NDA_W}) are
O(1). This gives us the following effective superpotential:\footnote{
For the K\"{a}hler potential, we assume the minimal form 
$\tilde{K} = \phi^{\dagger}\phi+\bar{\phi}^{\dagger}\bar{\phi}$, 
which leads to $K_{\rm comp} = \Phi^{\dagger}\Phi+
\bar{\Phi}^{\dagger}\bar{\Phi}$.}
%We assume that this can be done effectively by making a use of 
%a pair of composite chiral superfields 
%$\Phi$, $\bar{\Phi}$ which consist of $Q$ superfields and have U(1)$_m$ 
%charges $\pm 1$, respectively.
%The effective superpotential concerning $\Phi$, $\bar{\Phi}$, $Z$ 
%fields are given by 
%
\begin{eqnarray}
 W_{\rm comp} \simeq \Lambda \Phi \bar{\Phi} 
                 + \lambda Z \Phi \bar{\Phi}.
\end{eqnarray}
Then, the $m_E^2$ and $m_{\bar{E}}^2$ can be evaluated perturbatively 
as long as $\lambda \lsim O(1)$. 
The two-loop diagrams relevant to these masses 
are shown in Fig.~\ref{messenger_soft_mass}.
The resulting $m_E^2$ and $m_{\bar{E}}^2$ are positive and 
their order of magnitudes are given by
\begin{eqnarray}
 m_E = m_{\bar{E}} 
     \sim \frac{\alpha_m}{4\pi}\frac{\lambda F_Z}{\Lambda} 
     \simeq \frac{\alpha_m}{4\pi}\frac{\lambda^2}{16\pi^2}\Lambda,
\end{eqnarray}
where $\alpha_m = g_m^2/4\pi$ and $g_m$ is the U(1)$_m$ gauge coupling 
constant. We note that this result is also obtained by applying directly 
the NDA to $Q$-integration without the introduction of the composite
chiral superfields.

Unlike in the original model \cite{DNS,DNNS}, 
the SUSY-breaking effects are not 
transmitted to the messenger quarks and leptons with 
these positive soft SUSY-breaking masses.
Thus, the squarks, sleptons, and gauginos in the 
minimal supersymmetric SM (MSSM) sector remain massless at this level.

Now, we discuss transmission of the DSB effects to the MSSM sector.
For the positive $m_E^2$ and $m_{\bar{E}}^2$, 
the $S$ and $F_S$ do not develop vacuum-expectation values. 
In the present model, however, the $S$ field gets a negative 
soft SUSY-breaking mass squared $-m_S^2$ from the 'running effects' 
through loops of the $E$ and $\bar{E}$ fields with the Yukawa coupling 
$k_E S E \bar{E}$ in Eq.~(\ref{superpotential}).\footnote{
This possibility has been also noted in the original model \cite{DNS}.}
The $m_S^2$ is evaluated as 
\begin{eqnarray}
 m_S^2 \simeq \frac{4}{(4\pi)^2}k_E^2 m_E^2 \ln\frac{\Lambda}{m_E}. 
\end{eqnarray}
Altogether, the scalar potential for the messenger sector becomes
\begin{eqnarray}
 V_{\rm mess} &=& |k_E E \bar{E} + f S^2 + k_d d \bar{d} + k_l l \bar{l}|^2
          \nonumber\\ 
          &&  + |k_E S E|^2 + |k_E S \bar{E}|^2 \nonumber\\
          &&  + |k_d S d|^2 + |k_d S \bar{d}|^2 \nonumber\\
          &&  + |k_l S l|^2 + |k_l S \bar{l}|^2 \nonumber\\
          &&  + m_E^2 |E|^2 + m_E^2 |\bar{E}|^2 - m_S^2 |S|^2.
\end{eqnarray}
This potential has a global minimum at 
\begin{eqnarray}
 && \langle S^* S \rangle = \frac{m_S^2}{2 f^2}, \nonumber\\
 && \langle E \rangle = \langle \bar{E} \rangle =
 \langle d \rangle = \langle \bar{d} \rangle =
 \langle l \rangle = \langle \bar{l} \rangle = 0, \nonumber\\
 && \langle |F_S| \rangle = \frac{m_S^2}{2 f},
\end{eqnarray}
in a certain parameter region.
Thus, all the SM gauge symmetries are preserved and 
the SUSY-breaking effects are transmitted to the messenger quark and 
lepton multiplets through $\langle F_S \rangle$.\footnote{
The messenger gauge group U(1)$_m$ is also unbroken in this vacuum.
Although the massless U(1)$_m$ gauge boson couples to the 
SM quarks and leptons through higher loops, we find that 
this causes no phenomenological problem, since induced forces are short
range and very weak. 
We should note also that there are stable U(1)$_m$ charged particles 
in the present model. All such particles, however, have the 
masses of order $\Lambda \simeq 10^{9-10}~\rm{GeV}$. 
Thus, the cosmological density of these particles is expected to be 
very low since the reheating temperature $T_R$ of the inflation 
should be $T_R \lsim 10^{6}~\rm{GeV}$ \cite{MMY} to solve 
the gravitino problem in the present model, and hence they cause no 
phenomenological problem.}

When we integrate out the messenger quarks and leptons 
$d$, $\bar{d}$, $l$, and $\bar{l}$, 
the MSSM gauginos acquire masses at the one-loop level as 
\begin{eqnarray}
 m_{\tilde{g}_i} = c_i \frac{\alpha_i}{4\pi} \Lambda_{\rm mess}, 
\end{eqnarray}
where $c_1 = 5/3$, $c_2 = c_3 = 1$, and $m_{\tilde{g}_i}$ ($i=1, \cdots, 3$) 
denote the bino, wino, and gluino masses, respectively.
The soft SUSY-breaking masses for squarks, sleptons, and Higgses 
$\tilde{f}$ in the MSSM sector are generated at the two-loop level as 
\begin{eqnarray}
 m_{\tilde{f}}^2 = 2 \Lambda_{\rm mess}^2 \left[ 
                   C_3 (\frac{\alpha_3}{4\pi})^2 + 
                   C_2 (\frac{\alpha_2}{4\pi})^2 + 
                   \frac53 Y^2 (\frac{\alpha_1}{4\pi})^2  \right], 
\end{eqnarray}
where $C_3 = 4/3$ for color triplets and zero for singlets, 
$C_2 = 3/4$ for weak doublets and zero for singlets, 
and $Y$ is the SM hypercharge ($Y = Q_{\rm em} - T_3$).
Here, $\Lambda_{\rm mess}$ is an effective messenger scale defined as 
\begin{eqnarray}
 \Lambda_{\rm mess} \equiv \frac{\langle |F_S| \rangle}{\langle S \rangle} 
               = \frac{m_S}{\sqrt{2}}, 
\end{eqnarray}
which can be written in terms of the SUSY-breaking scale 
$\sqrt{F_Z}$ as 
\begin{eqnarray}
 \Lambda_{\rm mess} &\simeq& \frac{\sqrt{2}}{(4\pi)^4} \alpha_m 
           \lambda^2 k_E \sqrt{\ln \frac{(4\pi)^3}{\alpha_m \lambda^2}} 
           \cdot \Lambda \nonumber\\
 &=& \frac{\sqrt{2}}{(4\pi)^4} \alpha_m \lambda \sqrt{\lambda} k_E 
           \sqrt{\ln \frac{(4\pi)^3}{\alpha_m \lambda^2}} 
           \cdot \sqrt{F_Z}.
\end{eqnarray}
If we intend to make the MSSM gaugino and sfermion masses 
at the electroweak scale, the effective messenger scale 
$\Lambda_{\rm mess}$ must be $10^{4-5}~\rm{GeV}$.
Then, the SUSY-breaking scale $\sqrt{F_Z}$ becomes 
$10^{8-9}~\rm{GeV}$ for $\alpha_m \simeq 10^{-2}$ 
and $\lambda \sim k_E \sim O(1)$.\footnote{
In order that the U(1)$_m$ gauge coupling constant should not blow up 
below the grand unification scale $\simeq 
2 \times 10^{16}~\rm{GeV}$, $\alpha_m$ at the $\Lambda_{\rm mess}$ 
must be smaller than $\sim 3 \times 10^{-2}$ \cite{GMM}.
A similar requirement for the Yukawa coupling constants 
$\lambda$ and $k_E$ leads to 
$\lambda \lsim 2$ and $k_E \lsim 1$ at the $\Lambda_{\rm mess}$.}
This corresponds to the gravitino mass, 
\begin{eqnarray}
 m_{3/2} \simeq \frac{F_Z}{\sqrt{3}M_G} \simeq 10~\rm{MeV} - 1~\rm{GeV},
\end{eqnarray}
where $M_G$ is the gravitational scale $M_G \simeq 2.4 
\times 10^{18}~\rm{GeV}$.
Thus, dangerous flavor-changing neutral currents are suppressed 
well below the present experimental limits \cite{GGMS}.

Finally, we should stress that the dangerous D-term problem 
\cite{DNS} for the U(1)$_m$ is automatically solved in the present model.
That is, there is a charge conjugation symmetry 
in the DSB sector defined as 
\begin{equation}
\begin{array}{c}
 Q_1 \leftrightarrow Q_2, \quad
 V \leftrightarrow -V,\nonumber\\
\\
 Z^{12} \leftrightarrow -Z^{12}, \quad 
 Z^{13} \leftrightarrow Z^{23}, \quad 
 Z^{14} \leftrightarrow Z^{24}, 
\end{array}
\end{equation}
which forbids the Fayet-Iliopoulos D-term of $V$ \cite{DNNS,FNPRS}.
Here, $V$ is the U(1)$_m$ gauge superfield.
This may be a big merit of the present vector-like 
gauge theory \cite{IY,IT}.

In conclusion, we have constructed a gauge mediation model 
which has a unique color-preserving global minimum 
in a certain parameter region.\footnote{
It may not be excluded that the original model \cite{DNS,DNNS} 
has also such a desired global minimum in the nonperturbative regime.}
The SUSY-breaking scale turns out to be as high as 
$\langle \sqrt{F_Z} \rangle \sim 10^{8-9}~\rm{GeV}$ 
which results in $m_{3/2} \simeq 10~\rm{MeV} - 1~\rm{GeV}$.
The present model is perfectly natural and phenomenologically consistent.
However, it should be noted that if there exist moduli fields 
like a dilaton in superstring theories with masses 
$\sim m_{3/2}$, the present model has a serious cosmological 
problem, since decays of such moduli fields produce too much background 
$\gamma$ rays as pointed out in the recent work \cite{HKY}.
However, if the presence of light moduli fields is an inevitable 
prediction of the superstring theories is not yet clear to us.\\

One of the authors (K.T.) would like to thank T. Moroi
for useful comments.
The work of K.T. is supported by DOE contract DOE/ER/01545-726.

\newpage

%
%%%%%%%%%%%%%%%%%%%%%%%%%%%%%%%%%%%%%%%%%%%%%%%%%%%%%%%%%%%%%%%
%
% NEW COMMANDS FOR THE BIBLIOGRAPHY
%
%%%%%%%%%%%%%%%%%%%%%%%%%%%%%%%%%%%%%%%%%%%%%%%%%%%%%%%%%%%%%%%
\newcommand{\Journal}[4]{{\sl #1} {\bf #2} {(#3)} {#4}}
\newcommand{\APJ}{Ap. J.}
\newcommand{\CJP}{Can. J. Phys.}
\newcommand{\NC}{Nuovo Cimento}
\newcommand{\NP}{Nucl. Phys.}
\newcommand{\PL}{Phys. Lett.}
\newcommand{\PR}{Phys. Rev.}
\newcommand{\PRep}{Phys. Rep.}
\newcommand{\PRL}{Phys. Rev. Lett.}
\newcommand{\PTP}{Prog. Theor. Phys.}
\newcommand{\SJNP}{Sov. J. Nucl. Phys.}
\newcommand{\ZP}{Z. Phys.}
%%%%%%%%%%%%%%%%%%%%%%%%%%%%%%%%%%%%%%%%%%%%%%%%%%%%%%%%%%%%%%%

%
\end{document}